# Combining incentives for pollination with collective action to provide a bundle of ES in farmland

Jérôme Faure[1,2,4], Lauriane Mouysset[4], Sabrina Gaba[2,3]

Affiliations:

[1] CNRS, UMR 5113, GREThA, Université de Bordeaux, avenue Léon Duguit, 33608, Pessac cedex, France

[2] INRAE, USC 1339, Centre d'Études Biologiques de Chizé, F-79360 Villiers-en-Bois, France

[3] CNRS & Université de La Rochelle, UMR 7372, Centre d'Études Biologiques de Chizé, F-79360 Villiers-en-Bois, France

[4] Centre International de Recherche sur l'Environnement et le Développement (CIRED), 45bis Avenue de la Belle Gabrielle, 94130 Nogent-sur-Marne, France

*Corresponding author:* Jerome Faure,

Centre International de Recherche sur l'Environnement et le Développement (CIRED), 45bis Avenue de la Belle Gabrielle, 94130 Nogent-sur-Marne, France,

jerome.faure@cebc.cnrs.fr

*Running head*: Incentives and collective action

*Pages: 50 (63 with appendices)*

# Abstract

A polycentric approach to ecosystem service (ES) governance that combines individual incentives for interdependent ES providers with collective action is a promising lever to overcome the decline in ES and generate win-win solutions in agricultural landscapes. In this study, we explored the effectiveness of such an approach by focusing on incentives for managed pollination targeting either beekeepers or farmers who were either in communication with each other or not. We used a stylized bioeconomic model to simulate (i) the mutual interdependency through pollination in intensive agricultural landscapes and (ii) the economic and ecological impacts of introducing two beekeeping subsidies and one pesticide tax. The findings showed that incentives generated a spillover effect, affecting not only targeted stakeholders but non-targeted stakeholders as well as the landscape, and that this effect was amplified by communication. However, none of the simulated types of polycentric ES governance proved sustainable overall: subsidies showed excellent economic but low environmental performance, while the tax led to economic losses but was beneficial for the landscape. Based on these results, we identified three conditions for sustainable ES governance based on communication between stakeholders and incentives: (i) strong mutual interdependency (i.e. few alternatives exist for stakeholders), (ii) the benefits of communication outweigh the costs, and (iii) the incentivized ES drivers are not detrimental to other ES. Further research is needed to systematize which combination of individual payments and collaboration are sustainable in which conditions.

# Key words

# Introduction

Agricultural landscapes represent more than one-third of viable lands used for crop production or livestock (FAO, 2021). Although these productive socio-ecological systems mainly aim to deliver a sole ecosystem service (ES) – i.e. the production of food and/or feed – agricultural landscapes remain important providers of a range of ES (Swinton et al., 2007). Yet management interventions such as changes in land use or fertilizer or pesticide use can affect one ES or multiple ES simultaneously (Bennett et al., 2009). Currently, intensive agricultural practices are negatively impacting ecosystem functioning and ES provision (Diaz et al., 2019; Power, 2010). Schemes such as payments for ecosystem services (PES) schemes and, more generally, individual monetary incentives have been one policy response to halt this loss in ES; this is the case in Europe as set out in the second pillar of the Common Agricultural Policy on rural development and good practice (CAP; Simoncini et al., 2019). However, this 'top-down' approach to ES governance has been widely criticized (Muradian, 2013; Pe'er et al., 2020), mainly because it simplifies ES governance by targeting a unique stakeholder, while ES are provided to and/or benefit many stakeholders, resulting in the under-provision of ES and local conflicts (Muradian et al., 2010; Barnaud et al., 2018). Polycentric ES governance that relies on collective action and market-based incentives has been advocated as an alternative more suited to the nature of ES as a public good (Öström, 2010a; Muradian et al., 2012). It has been argued that a combination of top-down instruments and bottom-up initiatives could more effectively increase the supply of ES and thus satisfy ES demand from multiple stakeholders (OECD, 2013; Prager, 2015; Salliou et al., 2017). However, to make such an approach operational for implementation in post-2020 CAP goals, for instance, it is necessary to identify how to optimally combine monetary incentives and collective action to ensure both effective and sustainable ES governance.

Collective action can be defined as an *action taken by a group in pursuit of members' perceived shared interests* (Scott and Marshall, 2009). Different modes of collective action exist in agricultural landscapes, ranging from coordination to collaboration. Following Prager (2015), we define collaboration as a process in which stakeholders in land management actively meet, work and talk together. We differentiate this from coordination, in which stakeholders work individually towards a same objective. Some PES schemes such as collective payments already rely on collaboration (Kerr et al., 2014): stakeholders are paid for a common result. One difficulty in implementing collective payments is that they may involve real or perceived free riding and create issues of trust or be incompatible with farm management scale (OECD, 2013; Prager, 2015). Other forms of collaboration between stakeholders include dialogue and communication based on shared information about economic strategy (Chwe, 2000; Opdam et al., 2016). These forms are typically easier to initiate and more compatible with individual incentives (OECD, 2013).

In agricultural landscapes, willingness to act collectively increases when stakeholders are mutually interdependent (Barnaud et al., 2018): that is, they feel they need one another to solve a problem or improve their situation. This creates a symmetric relationship, with the emergence of shared interests and reciprocity facilitating communication (Ostrōm, 1998; Vanni, 2013). These situations are common in agricultural landscapes, as ES providers are also ES beneficiaries, depending on services supplied by other ES providers (Vialatte et al., 2019). Thus, one might expect polycentric forms of governance that combine incentives and communication to be effective in such situations of mutual interdependency.

In this study, we explored polycentric governance of pollination ES in an agricultural landscape. Pollination is a critical ES for agriculture: 70% of the world's cultivated crops depend on insect pollination (Klein et al., 2007), a figure rising to more than 85% in Europe (Williams, 1994). For example, the most cultivated oilseed crop in Europe, oilseed rape (*Brassica napus L.*), relies on insect pollination for more than 35% of its production (Perrot et al., 2019; Woodcock et al., 2019). Pollination services have been valued at US$235–577 billion each year (Lautenbach et al., 2012). Both honeybees (*Apis mellifera*), managed by beekeepers, and unmanaged wild bees play a vital role in enhancing seed quantity and quality (Bommarco, Marini & Vaissière, 2012). Yet both domestic and wild bees worldwide are declining in agricultural landscapes, mainly due to intensive pesticide use and the decline in semi-natural habitats (Potts et al., 2016). This deepens the mismatch between pollination supply and demand, as farmers themselves reported in a recent survey at European scale (Breeze et al., 2019).

Farmers and beekeepers are mutually interdependent on pollination (it is their common interest): beekeepers benefit from farmers and vice versa (Bretagnolle and Gaba, 2015; Vialatte et al., 2019). By keeping honeybees, beekeepers are pollination providers. Through crop flower visitation, honeybees have been identified as the main pollinator for mass-flowering crops such as oilseed rape (Perrot et al., 2018) and sunflowers (Perrot et al., 2019), resulting in higher yields and incomes (Catarino et al., 2019). In return, mass-flowering crops provide abundant floral resources for honeybees, therefore ensuring honey production and overwinter survival success (Requier et al., 2015). In this way, farmers are also indirect providers of managed pollination. Consequently, farmers and beekeepers need one another to maximize their benefits – they are mutually interdependent. Their economic decisions, such as land and pesticide use or the number of beehives, may influence each other depending on their relationship. There are diverse types of beekeeper/farmer collaboration around the world, ranging from informal arrangements to pollination markets (Narjes et al., 2019). In Europe, collaboration is rather scarce, and if it exists it is mainly carried out through face-to-face communication (Breeze et al., 2019). This can be measured as the amount of intentionally shared information. For example, beekeepers may inform farmers about their intention to increase beehive capacity, which can impact farmers' land-use decisions. In this context, policies encouraging managed pollination (such as beekeeping subsidies) or discouraging detrimental factors (such as pesticide taxes), may lead to economic spillovers (i.e. indirect economic effects on non-targeted stakeholders) that are conditioned by communication. Beyond economic effects on stakeholders, trade-offs between ES or their drivers may also lead to environmental impacts at the landscape scale (e.g. by reducing pesticide use; Bennett et al., 2009).

In this study, we explicitly considered the mutual interdependency between farmers and beekeepers and their level of communication using a bioeconomic modelling approach. We examined whether incentives motivating one of these ES providers to change ES drivers could lead to the delivery of a bundle of ES, therefore benefiting other stakeholders as well as the landscape. Following recommendations from beekeepers (Breeze et al., 2019), we considered (i) subsidies targeting beekeepers, which seem effective for increasing the number of honeybees (Stefanic et al., 2004; Çevrimli, 2019) and (ii) a tax targeting farmers aimed at reducing the use of pesticides in rapeseed fields, i.e. a pesticide tax (Finger et al., 2017). We posited that more sustainable pollination governance (i.e. higher economic and environmental performance) can be achieved by combining the implementation of these incentives with enhanced communication between farmers and beekeepers. In our model, we assessed the performance of pollination incentives with or without communication from both an environmental and ecological perspective.

# Materials and methods

## Model overview and interdependency

We developed a bioeconomic model of an intensive agricultural landscape that included two economic activities, agriculture and beekeeping, following the framework of Bretagnolle and Gaba (2015). The economic decision-making model and the ecological model were linked by pollination. The economic model included farmers' strategies on land use and pesticide use, and beekeepers' strategies on the number of beehives. The ecological model predicted populations of honeybees and wild bees. An overview of the model is presented in Figure 1a, in which arrows represent the mathematical relationships (see below for details). The model also accounted for the mutual interdependency between farmers and beekeepers: farmers depend on beekeepers through pollination by domestic bees, whose abundance is defined by beekeepers by the number of beehives (Figure 1b), while beekeepers depend on farmers because mass-flowering crops provide important floral resources for domestic bees and because pesticide use directly affects bee survival (Figure 1b). Lastly, the model included whether or not farmers and beekeepers communicated, as this can affect their respective management decisions and hence the efficiency of domestic bee pollination.

**Agricultural production**

Three representative crops of intensive agricultural cultivation were modelled (these crops represent more than 80% of existing crops in intensive landscapes; Bretagnolle et al., 2018): wheat for winter cereals (subscript $W$), temporary grasslands used for hay for meadows (subscript $G$) and oilseed rape for oilseed crops (subscript $OSR$). Oilseed rape is largely dependent on pollination (Perrot et al., 2018) and is also the preferred crop of European beekeepers due to its high nectar capacity (Breeze et al., 2019). In the model, each farmer owned his/her farm area (normalized to 1 ha) and cultivated the three crops in the following proportions $x_W \in [0,1]$, $x_G \in [0,1]$ and $x_{OSR} \in [0,1]$ for wheat, grasslands and oilseed rape respectively. In the following, the vector of crop proportion is noted $X$, corresponding to the land use in the landscape. We imposed a constant acreage, such as $x_W + x_G + x_{OSR} = 1$. In oilseed rape fields, farmers applied pesticides at the rate $x_p \in [0,1]$. Pesticide use in other crops was assumed invariable. The representative farmer's decisions about land use and agrochemical use were made to maximize the profit function $\Pi^{(f)}$ (Figure 1a):

$$\max_{X,x_p}\{\Pi^{(f)}(X,x_p) = \sum_{k\in\{W,G\}} \eta_k(x_k) + p_{OSR}F_{OSR}(X,x_p,x_H) - C_{OSR}(x_{OSR},x_p) + \tau^{(f)}(X,x_p)\}$$

$$\text{Subject to} \begin{cases} x_W \in [0,1] \\ x_G \in [0,1] \\ x_{OSR} \in [0,1] \\ x_W + x_G + x_{OSR} = 1 \\ x_p \in [0,1] \end{cases} \quad (1)$$

$\eta_W$ and $\eta_G$ are the wheat and grassland benefit functions respectively presented in Table 1 by Eq.2 and Eq.3. $F_{OSR}$ is the OSR production function. $p_{OSR}$ is the sale price of OSR seeds, which was assumed to be constant and independent of farmers' decisions (i.e. the farmers are price-takers). The function $C_{OSR}$ represents OSR production costs, and was assumed to be linear with respect to the production factors $x_{OSR}$ and $x_p$. The function $\tau^{(f)}$ is the economic incentive aiming to support pollination (see '*Policy scenarios*' section below). The dependence of farmers on beekeepers is reflected by $x_H$, which is the number of beehives set up by the representative beekeeper (Figure 1b).

Following Montoya et al. (2019), oilseed rape yield was represented by three partially additive yields (Eq.4 in Table 1): a part dependent on both wild bees and honeybees $f_{OSR,1}$ (Eq.5 in Table 1), a part dependent on pesticide application $f_{OSR,2}$ (Eq.6 in Table 1), and a part independent of both bee abundance and pesticide application $f_{OSR,3}$. The partial yields $f_{OSR,1}$ (and $f_{OSR,2}$) follow a saturating, type II functional response that is increasing, concave. In the model, wheat plants did not depend on insect pollination, and we further assumed that hay production did not depend on pollination by bees.

## Honey production

The honey production of each beekeeper depended on the number of beehives set up, or $x_H$. We assumed that beekeepers' labour capacity was constant, with the number of beehives limited to $\overline{x_H}$ per beekeeper. The representative beekeeper chose $x_H$ beehives to maximize his/her profit $\Pi^{(k)}$ in function (Figure 1a):

$$\max_{x_H}\{\Pi^{(k)}(x_H, x_{OSR}, x_p) = p_H F_H(x_H, x_{OSR}, x_p) - C_H(x_H) + \tau^{(k)}(x_H, x_{OSR}, x_p)\}$$
$$\text{Subject to } \{x_H \in [0, \overline{x_H}]\} \quad (9)$$

where $F_H$ is the honey production function, $p_H$ the price of honey, and $C_H$ the cost function of the inputs, which were assumed to be linear with respect to the number of hives. $\tau^{(k)}$ is the economic incentive function and depends on the policy (see 'Policy scenarios' section below). Honey production followed a Cobb-Douglas function with $x_{OSR}$ and $x_H$ as inputs (Eq.7 in Table 1). To account for competition for floral resources between colonies (Champetier, Sumner & Wilen, 2015), the marginal productivity of beehives decreased (elasticity less than 1). The area of oilseed rape in the landscape was included in the production function as honey production varies widely with the amount of mass-flowering crops (Free, 1993). As has been empirically shown by Chambers et al. (2019), the use of pesticides implies losses in honey production due to honeybee mortality. We included this damage (i.e. the negative effect of pesticides) through $D \in [0,1]$ in the honey production function; Eq.7 and 8 in Table 1. The dependence of beekeepers on farmers is reflected by $x_{OSR}$, which is the main source of nectar for honeybees, and by $D$, which is the pesticide damage to honeybees (Figure 1b).

**Bee populations**

The honeybee population varied with the number of beehives ($x_H$). The parameter $k_H$ is the carrying capacity per hive, comprising a survival rate that was assumed to be constant (Eq.10). We assumed that there were no 'empty hives', a phenomenon reported by Stokstad (2007), which is already included in the carrying capacity of honeybees. Since the beehive number is derived from economic decision-making (Eq.9), it already includes the pesticide effect (Eq.7, Table 1). In accordance with Chambers et al. (2019), we computed and evaluated the rate between honey and domestic bee losses (Appendix S1).

Following Kleczkowski et al. (2017), Eq.10 shows that the wild bee population size depends on the area of grassland, the bees' carrying capacity $k_w$ and the use of pesticides (Appendix S1). This assumed constant nesting resources and no direct competition between wild and domestic bees, as reported by Lindström et al. (2016). The total population of bees is :

$$B(x_H, x_G, x_{OSR}, x_p) = \overbrace{x_H k_H}^{honeybees} + \overbrace{x_G k_w D(x_{OSR}, x_p)}^{wild\ bees} \quad (10)$$

**Policy scenarios for improving pollination**

In a recent survey, European beekeepers viewed financial incentives to encourage beekeeping activity (e.g. to increase the number of hives) and reduction of pesticide use as effective ways to support honeybees and enhance crop pollination ES (Breeze et al., 2019). We thus implemented three financial incentive scenarios aiming to increase pollination: two using beekeeping subsidies and one imposing an pesticide tax.

- Scenario HS: in this scenario a *per hive subsidy* (HS) was implemented, formulated as amount $\tau^{(k)} = z_{HS} x_H \geq 0$, which was granted to each beekeeper proportionally to the number of beehives set up. $z_{HS}$ is the sum granted per beehive.

- Scenario PS: in this scenario a *price subsidy* (PS) was implemented, formulated as amount $\tau^{(k)} = z_{PS} F_H \geq 0$, which was granted to the beekeeper for each kilogram of honey sold. $z_{PS}$ is the sum granted per kilogram of honey.

- Scenario PT: in this scenario a *pesticide tax* (PT) was implemented, formulated as amount $\tau^{(f)} = z_{PT} x_{OSR} x_p \leq 0$, which was levied on farmers according to the level of pesticides applied on their oilseed rape fields. $z_{PT}$ is the tax per hectare of oilseed rape.

These scenarios were compared to a *business as usual* (BAU) control scenario with no economic incentive ($\tau^{(f)} = \tau^{(k)} = 0$).

## ES governance effectiveness and bioeconomic indicators

We defined ES governance effectiveness as the simultaneous increase in pollination and economic gains. We assessed this effectiveness in scenarios with two economic indicators (total and stakeholder wealth) and with a pollination indicator (Table 2). In addition, we extended our analysis on ES governance sustainability using three other ES indicators (food provision, water quality concerning pesticides, and water quality concerning nutrients) and two indicators of biodiversity conservation (wild bee abundance and plant species richness). We also defined marginal indicators at the stakeholder scale to assess the direct effects and indirect (spillover) effects of the incentives. We calculated input marginal product (output mass per unit area of OSR and per pesticide unit for farmers, output mass per beehive for beekeepers) and output marginal revenue (€.ton$^{-1}$ OSR for farmers, €.kg$^{-1}$ honey for beekeepers) (see Appendix S2 for more details).

## Communication and decision-making

As beekeepers and farmers are mutually interdependent, their decisions are tightly linked; thus they can adopt a concerted strategy, i.e. a collective action. Feedback from the field showed that farmers and beekeepers sometimes communicate on their strategies, sharing their intentions and coordinating their actions together. For example, farmers may provide a location for hives next to oilseed rape fields or may reduce pesticide intensity or timing to decrease the negative impact on domestic bees. In exchange, beekeepers may set up more beehives than initially planned (Allier, 2012; Breeze et al., 2019). Yet this communication is not always easy, for reasons including a knowledge gap about their common interest (Breeze et al., 2019), social distance, or the lack of an institutional setting (Oström, 1990; Chwe, 2000).

In our model, the decision vector contained the values of the control variables ($x^*_{OSR} x^*_p\ x_W\ x_G$) for the representative farmer and $x^*_H$ for the representative beekeeper, based on maximizing their respective profits. The decision-making result depends on the communication level, which can be modelled as the quantity of information shared and thus 'owned' by the stakeholders. As described above, communication depends on specific criteria and can be difficult to initiate: in economic terms, communication involves transaction costs for stakeholders. We chose to model two highly contrasting transaction cost situations. The first was a situation in which (i) the costs of communication exceed the benefits so stakeholders do not share any information about their intention. For example, a situation in which farmers are not even aware of the presence of beehives close to their fields, as has been observed in some cases (Allier, 2012). The second was a situation in which (ii) communication has no cost so stakeholders have complete information regarding each other's intentions. For example, the farmers and beekeepers are friends and fully share their intentions. From a game theory perspective, these situations are coordination, non-cooperative games (i.e. separate and individual decision models; Matsumoto & Szidarovszky, 2016) (i) without and (ii) with complete information (Oliver, 1993).

We assumed that the BAU scenario decisions depended on the path: farmers and beekeepers have historically adapted their strategies relative to each other and thus are fully informed. Hence, only for incentive scenarios did we differentiate between a case in which the stakeholders (i) do not or (ii) do communicate their intentions. When they do not communicate, non-incentivized stakeholders ignore incentivized stakeholders' strategies. When they do communicate, the system evolves according to a

leader–follower sequential decision-making process, in which the representative farmer maximizes his/her profit first (leader) and the representative beekeeper accepts the farmer's decision (follower) (Matsumoto & Szidarovszky, 2016). We used backward induction to maximize their profits, using an analytical formulation of the beekeeper's optimum first. This analytical expression was then used in the profit maximization procedure for the farmers. The optimums were determined by numerical optimization using the *optim* function in the "stats" package in *R v3.4.4* (R Core Team 2018).

**Parameter calibration and sensitivity analysis**

The parameter values for oilseed rape production were estimated using a dataset from the Zone Atelier Plaine & Val de Sèvre Long-term Socio-ecological Research study area (Bretagnolle et al., 2018). For other parameters, we used values taken from relevant studies (Appendix S3). As suggested by Muradian and Rival (2012), we implemented realistic rather than Pigouvian levels of incentives (Pindyck and Rubinfeld, 2018). For the subsidy calibrations, the amount granted to each beekeeper was fixed at €5000, corresponding to a realistic budget for a professional beekeeper in France in 2020 (Ministry of Agriculture Food, 2020). To compare the two subsidy scenarios, subsidies were set so that the total amount allocated to the beekeeper was equal to €5000, allowing €27.hive$^{-1}$ for the hive subsidy and €1.10.kg$^{-1}$ for the honey price subsidy. The tax per hectare was fixed at €50.ha$^{-1}$ for the maximal agrochemical application rate ($x_p = 1$), in accordance with pesticide taxation literature (Jacquet, Butault & Guichard, 2011; Böcker & Finger, 2016). We defined the indicators at the landscape scale as follows: in Deux-Sèvres (France), the average area per farm is 95 ha and we assumed that the modelled landscape consisted of 570 ha of agricultural lands, with $n = 6$ farmers and $m = 1$ beekeeper. A sensitivity analysis was performed on the oilseed rape price (Appendix S4).

# Results

**Economic effectiveness and spillover effects**

In order to grasp if a combination of incentives and communication between farmers and beekeeper would result in economically effective ES governance (i.e. all stakeholders win to communicate), we studied the effect of the three defined scenarios on the economic performance of agriculture and beekeeping as well as on public finances (Figure 2). As expected, the latter decreased with hive and price subsidies (as subsidies generate public expenses, Figure 2a) and increased in the pesticide tax scenario (as taxes generate public gains, Figure 2b-c). However, the effect of public policies on the private sector was more complex. While the subsidy scenarios targeted only beekeeping, they also benefited farmers, revealing a spillover effect (Figures 2a-b). This was accentuated by communication between stakeholders: for example, communication increased agricultural wealth by more than €4000 with the hive subsidy – i.e. 48% more than without communication (Figure 2a). We also found an increase in the beekeeper's wealth when the stakeholders communicated, so communication seemed to be a win-win strategy. This suggests that a combination of subsidies and communication allow effective polycentric ES governance since it is in the interest of beekeepers and farmers to communicate, resulting in an overall economic increase. In contrast, in the tax scenario, which targeted only farmers, we observed a spillover effect that was negative for both the farmers and the beekeeper (Figure 2c). In this scenario, communication benefited beekeeping by buffering the losses (by 50%, Figure 2c), but was prejudicial for farmers, who incurred higher losses. Hence, the tax scenario was not economically effective.

**Bioeconomic performance of the three policy scenarios**

In order to assess if the incentives combined with communication would lead to environmentally sustainable ES governance, we also assessed the bioeconomic performance of the three policy scenarios. Compared to the BAU scenario, when stakeholders communicated (Figure 3a), only subsidy policies increased the pollination service, with the highest increase achieved by the hive subsidy scenario (+62%), followed by the honey price subsidy scenario (+26.8%). Surprisingly, the pesticide tax scenario actually decreased pollination (-7%). Also surprisingly, the increase in pollination services decreased the overall environmental conditions of the landscape in subsidy scenarios. The subsidies reduced the magnitude of water quality services by 4.6% and 3% for the pesticide indicator, and by 24.8% and 15.1% for the nutrient indicator. Subsidies also had a negative effect on wild bee abundance (-29.8% hive subsidy and -18.2% price subsidy) and plant species richness (-6.9% HS and -4% PS). In contrast, the pesticide tax increased water quality services (i.e. by 5.5% and 10.4% for the pesticide and nutrient indicators) as well as wild bee abundance (+15.3%) and plant species richness (+2.5%).

At landscape scale, only the subsidy scenarios generated economic gains (+3.7% HS and +3.2% PS in total wealth), contrary to the tax scenario, which slightly reduced total wealth (-0.8%) (Figure 3a). Furthermore, the tax scenario decreased the stakeholders' wealth (-1.28%), while subsidies increased it (+5% HS and +3% PS).

The absence or presence of communication between the stakeholders generated differing bioeconomic performance (Figure 3b). Without communication, the positive effects of pollination gains were lowered by 21.5% and 20% in hive and honey price subsidy scenarios, but increased from -7% to +7% in the pesticide tax scenario. Indicators related to production (food provision) and economic performance (stakeholders' and total wealth) were also lower in the absence of communication in both subsidies and the tax scenario. Benefits to total wealth decreased by 46% in hive and 40% in price subsidy scenarios. Conversely, environmental indicators (pesticide and nutrient pollution reduction, wild bee abundance and species richness) increased. Compared to the BAU scenario, subsidies had similar environmental performance for these indicators, whereas in the pesticide tax scenario, environmental performance increased by +60% in wild bee abundance and +23% in pesticide-related water quality.

To sum up, only the subsidy scenarios increased pollination services; however, these scenarios presented lower environmental performance than the pesticide tax. Conversely, the pesticide tax had negative economic impacts, while subsidies had high economic performance. Communication between farmers and beekeepers on their intended strategies alongside pollination policies had a positive effect on the policy's economic performance, but a negative impact on its environmental performance.

**Land use change**

In an additional step, we computed the land use change associated with all four scenarios including BAU (Figure 4). The optimal mix in the BAU scenario is 45.6% winter cereals (wheat), 28% grasslands and 26.4% oilseed crops. In the subsidy scenarios, the proportion of oilseed rape increased up to 46.4% with the hive subsidy and 36.3% with the price subsidy, while the proportion of winter cereals and grasslands decreased by up to 32.6% and 21.1% respectively with the hive subsidy and by up to 39.6% and 24.1% with the price subsidy. In the pesticide tax scenario, the proportion of oilseed rape decreased by up to 16.1%, while the proportion of wheat and grasslands increased by up to 53% and 30.9%. Our sensitivity analysis on the oilseed rape price showed that land use was widely dependent on price level (Appendix S4).

**Decisional cascade and agro-ecological interdependency**

Figure 5 illustrates the decisional cascade leading to the spillover effects and environmental impacts of the policies. The implementation of a particular subsidy or tax generates changes in stakeholders' decision-making that affect ecological and economic processes acting at the scale of the stakeholder and the landscape. An analysis of the decisional processes highlighted the crucial role of pollination as the keystone of the mutual interdependency between the two economic activities. The increase in beekeepers' revenues in the subsidy scenarios was generated by an increase in marginal revenue from honey (37% for HS and 17% for PS), which affected decisions about the number of beehives (Figure 5a). In turn, the increase in the number of beehives increased pollination (Figure 5b), which resulted in an increase in oilseed rape productivity of 5.9% for the hive subsidy and 2.8% for the price subsidy (Figure 4c). Conditional to communication between farmers and beekeepers and given this higher productivity, farmers increased the area of oilseed rape and decreased their use of pesticides with respect to BAU (Figure 5c). Due to farmer/beekeeper interdependency through pollination, farmers' decisions generated an increase in marginal revenue from honey, which translated into an increase in the number of beehives, which then benefited farmers, and so on. In the pesticide tax scenario, taxation decreased farmers' revenues and marginal revenue from oilseed rape (-2.3%) (Figure 5d). This caused farmers to modify their practices by reducing pesticide use on oilseed rape (-38.9%) and reducing the cultivated area of oilseed rape. The result was a landscape with lower pesticide damage (-4.4%) and higher pollination (+4.4%, Figure 5e). However, the lower pesticide use was not enough to counterbalance the reduction in floral resources (less mass-flowering crops) and so beehive productivity decreased by 10%. Conditional on communication, this led beekeepers to decrease their production by decreasing the number of beehives by 15% (Figure 5f), decreasing oilseed rape marginal revenue, and so on.

## Discussion

A polycentric ES governance approach that combines individual incentives targeting interdependent ES providers with communication that allows them to act collectively is a promising lever to overcome the decline in ES and generate win-win solutions in agricultural landscapes. This study explored this approach by focusing on incentives for pollination that targeted either beekeepers or farmers with and without communication between them. We used a stylized bioeconomic model to simulate the economic and ecological impacts of two types of beekeeping subsidies and one pesticide tax aimed at fostering managed pollination. The findings indicated that none of the simulated types of polycentric ES governance were sustainable overall: beekeeping subsidies showed excellent economic performance but low environmental performance, while the pesticide tax was beneficial for the landscape but led to economic losses.

A key highlight is that a combination of beekeeping subsidies and communication was more effective than subsidies on their own to support pollination and the economy. Another finding was that public expenditure was also largely exceeded by private benefits, implying a cost-effective policy (OECD, 2013). These results are consistent with Breeze et al. (2019), who assumed that communication benefits both beekeepers and farmers. Our results are also in line with the study by Opdam et al. (2016) that showed that ES provided at landscape scale (such as pollination) were improved by collaboration between multiple land users. Subsidies encourage beekeepers to increase their beehive capacity. This in turn increases OSR yield, increasing production and creating additional revenue for farmers. This spillover effect arises whatever the communication level and is inherent to any positive externality (Pindyck and Rubinfeld, 2018). However, our simulations showed that if beekeepers communicated about their new beehive strategy, farmers increased the area of OSR and decreased their pesticide use per hectare. This strategy has also been observed in oilseed rape grown for seeds (Allier, 2012). Consequently, the new farming strategy benefited bees, pollination and beekeepers. In turn, the latter decided to further increase beehive capacity. In this way, communication increased production capacity, a benefit of collective action that has often been reported (OECD, 2013). Communication between farmers and beekeepers was beneficial due to the reciprocal nature of their interdependency. If dependency is one-sided – that is, if the beekeeper is independent from the farmer for his/her

production – communication would not create additional gains as no gainful collaborative strategy would be possible. As highlighted by Barnaud et al. (2018), power symmetry and the existence of a shared economic interest are factors that promote successful collective action. Nevertheless, collective action can be advantageous even without pecuniary benefits, as stakeholders may not be solely motivated by profit (Ostrom, 2010b).

Beyond pollination, these mechanisms could be generalized to study other regulating ES that underlie mutual interdependency between stakeholders (e.g. biological control, water quality). In the case of biological control for example, incentives for semi-natural habitats combined with more communication between farmers could lead to more effective ES governance (Salliou et al., 2019; Opdam et al., 2016).

Conversely, the pesticide tax scenario led to losses for both types of stakeholder, indicating ineffective ES governance whatever the communication level. The resulting change in the farmers' strategy (mainly a reduction in the area of OSR) led to a decrease in floral resources for honeybees. This type of behaviour has been observed after restrictions on neonicotinoid use (Zhang et al., 2017). When farmers communicated their decision to reduce OSR cultivation, beekeepers changed their strategy by lowering the number of hives, resulting in lower economic performance. Two factors underlie the failure of the tax policy: first, the benefits caused by pesticide reduction (lower bee mortality and thus higher pollination) were not high enough to counterbalance the cost of paying the tax. Second, farmers had an alternative to avoid the tax (planting more winter cereals and grasslands), which was detrimental for beekeepers because these crops flower less. These results reveal that to be economically effective, polycentric governance combining incentives and communication (compared to incentives only) requires (i) strong mutual interdependency between stakeholders, with no or few alternatives allowing them to reach their goal another way, and (ii) that the benefits of communication outweigh their costs.

Our results also found that environmental performance was high with the tax policy, but low for subsidies. Water quality, species richness and wild bee abundance increased in the pesticide tax scenario, while they decreased in both beekeeping subsidy scenarios as the increase in oilseed rape area was mainly at the expense of grasslands. Such substitution is often observed in intensive agricultural landscapes (Bretagnolle and Gaba, 2021). The lower biodiversity (of plants and wild pollinators) in the subsidy scenarios is therefore not surprising, as plant species richness is generally higher in grasslands (Öckinger & Smith, 2007) and haylands (Gaba et al., 2020), and wild bees require adequate semi-natural habitats (here modelled by grasslands) for nesting and foraging (Bretagnolle & Gaba, 2015). Moreover, the decrease in wild bee abundance resulting from less grassland can be accentuated by competition between wild and managed pollinators (Lindström et al., 2016). While we did not model direct competition, our results highlighted indirect competition between wild and domestic bees and thus an antagonism between managed and unmanaged pollination ES. To ensure the delivery of pollination ES, a decrease in wild pollinators can be compensated to a certain extent by managed pollinators: availability of managed bees can mask a decrease in wild pollination services that would otherwise encourage farmers to adopt conservation measures (Kleczkowski et al., 2017). In this case, communication between farmers and beekeepers accentuated this replacement of wild pollination by managed pollination, further impacting the environmental performance. Lastly, we found that environmentally effective polycentric governance combining incentives and communication was characterized by incentivized ES drivers that were not detrimental to other ES. This is important as studies have shown that ES-oriented policies can lead to trade-offs in which the provision of one service increases while another declines (Bennett et al., 2009). This raises an additional criterion to be met for polycentric governance to be sustainable: the strategies chosen to deliver an ES targeted by the incentives should account for trade-offs between ES to avoid negative environmental, economic and societal impacts.

In this study, the form of collective action we considered was communication between farmers and beekeepers. Other collaborative actions exist, such as collective payments, for example, but feedback from the field showed that collaboration between farmers and beekeepers in Europe usually takes the form of informal arrangements, although few studies have been carried out on this topic. However, it may be valuable to explore other types of collaborative action: for instance, collective incentives to promote pollination (Prager et al., 2012). We modelled communication either with or without infinite transaction costs. However, in real life there is typically a mix of the two. Transaction costs of communication between beekeepers and farmers can arise from social distance (e.g. they do not have the same social networks) or from more cognitive factors such as beliefs or status quo biases (Breeze et al., 2019). In Europe, relationships between beekeepers and farmers have been degraded by recent debates on pesticide use, for example. Risk can also be a barrier: oilseed rape is largely avoided by beekeepers because of the perceived pesticide risk (Breeze et al., 2019). A second threat is default risk, which is initially large but decreases with repetition and trust (Ostrom, 2010b). More generally, ES are characterized by high uncertainty, or asymmetric information, which increases reticence to collaborate (Muradian, 2013). Uncertainty reflects the 'noise' underlying biophysical processes and ecological functions (e.g. unstable pollination, Garibaldi et al., 2011), but also socio-economic uncertainties underlying negotiations and markets (Ostrom, 2010b; Muradian, 2013). Including these transaction costs in the model may result in a reduction in the benefits of communication and thus lessen stakeholders' motivation to share information, increasing the realism of the simulation (Olson, 2009). The model also did not simulate the decision to communicate per se, but only explored two extreme cases. It would be valuable in future studies to include communication in the decision model. The likelihood to act collectively is a rich research field: studies have shown that this decision can be shaped by reputation, trust or reciprocity (Ostrom, 2010b). Our model did not explore the causal relationship between incentives and communication, but it has been shown that incentives can raise stakeholder awareness about their interdependency and encourage them to collaborate more (Barnaud et al., 2018; OECD, 2013).

*Conclusion*

This study quantified the benefits of certain policy measures and communication between interdependent stakeholders in agricultural landscapes. Driver-oriented incentives for individual ES can be an interesting tool for providers and beneficiaries of ES, and also allow the opportunity to act collectively to get the best out of such measures. In the case of pollination, this opens avenues for future collaboration between farmers and beekeepers. However, the sustainability of polycentric ES governance based on incentives and communication appears to depend on the type of implemented policy. We identified three conditions for sustainable ES governance: (i) strong mutual interdependency (and few alternatives for stakeholders), (ii) that the benefits of communication outweigh the costs, and (iii) that the incentivized ES drivers do not conflict with the delivery of other ES. The simulated scenarios of managed pollination policy did not meet these conditions, but other ES such as biological control may be more in line with them. Further research is therefore needed to design sustainable polycentric governance that accounts for the complex interactions between ES providers, and between ES providers and the environment. Engaging a variety of stakeholders to share interest in a policy is inherently challenging and will require support to facilitate collective action.

## Authors' contributions

JF, SG and LM came up with the concept and designed the methodology; JF wrote the first draft of the manuscript. All authors contributed critically to the drafts and gave final approval for publication.

## Declaration of competing interest

The authors declare that they have no known competing financial interests or personal relationships that could influence the work reported in this paper.

## Acknowledgements

This project was supported by the French Ministry of Ecology project 'Pollinators' (2017–2020) and the ANR IMAgHO project 'Increase Multifunctionality in Agroecosystems through Trophic Networks' (ANR-18-CE32-002). The study also received funding from the European Union's Horizon 2020 research and innovation programme under grant agreement SHOWCASE No. 862480. JF was supported by a PhD grant from the French Ministry of Research. SG and LM are funded by INRAE and CNRS, respectively.

# References

Allier, F., 2012. Pollinisation en production de semences oléagineuses : une coopération technique entre agriculteurs et apiculteurs (Cahier technique). ITSAP, Paris.

Barnaud, C., Corbera, E., Muradian, R., Salliou, N., Sirami, C., Vialatte, A., Choisis, J.-P., Dendoncker, N., Mathevet, R., Moreau, C., Reyes-García, V., Boada, M., Deconchat, M., Cibien, C., Garnier, S., Maneja, R., Antona, M., 2018. Ecosystem services, social interdependencies, and collective action: a conceptual framework. E&S 23, art15. https://doi.org/10.5751/ES-09848-230115

Bennett, E.M., Peterson, G.D., Gordon, L.J., 2009. Understanding relationships among multiple ecosystem services. Ecology Letters 12, 1394–1404. https://doi.org/10.1111/j.1461-0248.2009.01387.x

Böcker, T., Finger, R., 2016. European Pesticide Tax Schemes in Comparison: An Analysis of Experiences and Developments. Sustainability 8, 378. https://doi.org/10.3390/su8040378

Bommarco, R., Marini, L., Vaissière, B.E., 2012. Insect pollination enhances seed yield, quality, and market value in oilseed rape. Oecologia 169, 1025–1032. https://doi.org/10.1007/s00442-012-2271-6

Breeze, T.D., Boreux, V., Cole, L., Dicks, L., Klein, A.-M., Pufal, G., Balzan, M.V., Bevk, D., Bortolotti, L., Petanidou, T., Mand, M., Pinto, M.A., Scheper, J., Stanisavljević, L., Stavrinides, M.C., Tscheulin, T., Varnava, A., Kleijn, D., 2019. Linking farmer and beekeeper preferences with ecological knowledge to improve crop pollination. People and Nature 1, 562–572. https://doi.org/10.1002/pan3.10055

Bretagnolle, V., Berthet, E., Gross, N., Gauffre, B., Plumejeaud, C., Houte, S., Badenhausser, I., Monceau, K., Allier, F., Monestiez, P., Gaba, S., 2018. Towards sustainable and multifunctional agriculture in farmland landscapes: Lessons from the integrative approach of a French LTSER platform. Science of The Total Environment 627, 822–834. https://doi.org/10.1016/j.scitotenv.2018.01.142

Bretagnolle, V., Gaba, S., 2021. Designing Multifunctional and Resilient Agricultural Landscapes: Lessons from Long-Term Monitoring of Biodiversity and Land Use. Springer Nature Switzerland AG.

Bretagnolle, V., Gaba, S., 2015. Weeds for bees? A review. Agronomy for Sustainable Development 35, 891–909. https://doi.org/10.1007/s13593-015-0302-5

Catarino, R., Bretagnolle, V., Perrot, T., Vialloux, F., Gaba, S., 2019. Bee pollination outperforms pesticides for oilseed crop production and profitability. Proceedings of the Royal Society B: Biological Sciences 286, 20191550. https://doi.org/10.1098/rspb.2019.1550

Çevrimli, M., 2019. Assessment of the effects of subsidies to the beekeeping sector in Turkey on the number of hives and amount of honey produced. Veteriner Hekimler Derneği Dergisi 90, 115–121. https://doi.org/10.33188/vetheder.506916

Chambers, R.G., Chatzimichael, K., Tzouvelekas, V., 2019. Sub-lethal concentrations of neonicotinoid insecticides at the field level affect negatively honey yield: Evidence from a 6-year survey of Greek apiaries. PLoS One 14. https://doi.org/10.1371/journal.pone.0215363

Champetier, A., Sumner, D.A., Wilen, J.E., 2015. The Bioeconomics of Honey Bees and Pollination. Environ Resource Econ 60, 143–164. https://doi.org/10.1007/s10640-014-9761-4

Chwe, M.S.-Y., 2000. Communication and Coordination in Social Networks. The Review of Economic Studies 67, 1–16. https://doi.org/10.1111/1467-937X.00118

Díaz, S., Settele, J., Brondízio, E.S., Ngo, H.T., Guèze, M., Agard, J., Arneth, A., Balvanera, P., Brauman, K., Butchart, S.H., 2019. Summary for policymakers of the global assessment report on biodiversity and ecosystem services of the Intergovernmental Science-Policy Platform on Biodiversity and Ecosystem Services. Intergovernmental Science-Policy Platform on Biodiversity and Ecosystem Services.

FAO, 2021. World Agriculture: Towards 2015/2030 - An FAO perspective [WWW Document]. fao.org. URL http://www.fao.org/3/y4252e/y4252e06.htm (accessed 3.1.21).

Finger, R., Möhring, N., Dalhaus, T., Böcker, T., 2017. Revisiting Pesticide Taxation Schemes. Ecological Economics 134, 263–266. https://doi.org/10.1016/j.ecolecon.2016.12.001

Free, J.B., 1993. Insect pollination of crops. Insect pollination of crops.

Gaba, S., Chevirón, N., Perrot, T., Piutti, S., Gautier, J.-L., Bretagnolle, V., 2020. Weeds Enhance Multifunctionality in Arable Lands in South-West of France. Front. Sustain. Food Syst. 4. https://doi.org/10.3389/fsufs.2020.00071

Jacquet, F., Butault, J.-P., Guichard, L., 2011. An economic analysis of the possibility of reducing pesticides in French field crops. Ecological Economics, Special Section - Governing the Commons: Learning from Field and Laboratory Experiments 70, 1638–1648. https://doi.org/10.1016/j.ecolecon.2011.04.003

Kerr, J., Vardhan, M., Jindal, R., 2014. Incentives, conditionality and collective action in payment for environmental services. International Journal of the Commons 8, 595–616. https://doi.org/10.18352/ijc.438

Kleczkowski, A., Ellis, C., Hanley, N., Goulson, D., 2017. Pesticides and bees: Ecological-economic modelling of bee populations on farmland. Ecological Modelling 360, 53–62. https://doi.org/10.1016/j.ecolmodel.2017.06.008

Klein, A.-M., Vaissiere, B.E., Cane, J.H., Steffan-Dewenter, I., Cunningham, S.A., Kremen, C., Tscharntke, T., 2007. Importance of pollinators in changing landscapes for world crops. Proceedings of the Royal Society B: Biological Sciences 274, 303–313. https://doi.org/10.1098/rspb.2006.3721

Lautenbach, S., Seppelt, R., Liebscher, J., Dormann, C.F., 2012. Spatial and Temporal Trends of Global Pollination Benefit. PLOS ONE 7, e35954. https://doi.org/10.1371/journal.pone.0035954

Lindström, S.A.M., Herbertsson, L., Rundlöf, M., Bommarco, R., Smith, H.G., 2016. Experimental evidence that honeybees depress wild insect densities in a flowering crop. Proceedings of the Royal Society B: Biological Sciences 283, 20161641. https://doi.org/10.1098/rspb.2016.1641

Matsumoto, A., Szidarovszky, F., 2016. Game Theory and Its Applications. Springer Japan.

Ministry of Agriculture Food, 2020. Repeuplement du Cheptel | FranceAgriMer - établissement national des produits de l'agriculture et de la mer [WWW Document]. FranceAgriMer. URL https://www.franceagrimer.fr/Autres-filieres/Apiculture/Accompagner/Dispositifs-par-filiere/Programme-apicole-europeen-PAE-2020-2022/Repeuplement-du-Cheptel (accessed 3.27.20).

Montoya, D., Haegeman, B., Gaba, S., Mazancourt, C. de, Bretagnolle, V., Loreau, M., 2019. Trade-offs in the provisioning and stability of ecosystem services in agroecosystems. Ecological Applications 29, e01853. https://doi.org/10.1002/eap.1853

Muradian, R., 2013. Payments for Ecosystem Services as Incentives for Collective Action. Society & Natural Resources 26, 1155–1169. https://doi.org/10.1080/08941920.2013.820816

Muradian, R., Corbera, E., Pascual, U., Kosoy, N., May, P.H., 2010. Reconciling theory and practice: An alternative conceptual framework for understanding payments for environmental services. Ecological Economics, Special Section - Payments for Environmental Services: Reconciling Theory and Practice 69, 1202–1208. https://doi.org/10.1016/j.ecolecon.2009.11.006

Muradian, R., Rival, L., 2012. Between markets and hierarchies: The challenge of governing ecosystem services. Ecosystem Services 1, 93–100. https://doi.org/10.1016/j.ecoser.2012.07.009

Narjes, M.E., Lippert, C., 2019. The Optimal Supply of Crop Pollination and Honey From Wild and Managed Bees: An Analytical Framework for Diverse Socio-Economic and Ecological Settings. Ecological Economics 157, 278–290. https://doi.org/10.1016/j.ecolecon.2018.11.018

Öckinger, E., Smith, H.G., 2007. Semi-natural grasslands as population sources for pollinating insects in agricultural landscapes. Journal of Applied Ecology 44, 50–59. https://doi.org/10.1111/j.1365-2664.2006.01250.x

OECD, 2013. Providing Agri-environmental Public Goods through Collective Action. OECD Publishing. https://dx.doi.org/10.1787/9789264197213-en

Oliver, P.E., 1993. Formal Models of Collective Action. Annual Review of Sociology 19, 271–300. https://doi.org/10.1146/annurev.so.19.080193.001415

Olson, M., 2009. The Logic of Collective Action: Public Goods and the Theory of Groups, Second Printing with a New Preface and Appendix. Harvard University Press.

Opdam, P., Coninx, I., Dewulf, A., Steingröver, E., Vos, C., van der Wal, M., 2016. Does information on landscape benefits influence collective action in landscape governance? Current Opinion in Environmental Sustainability, Sustainability governance and transformation 2016: Informational governance and environmental sustainability 18, 107–114. https://doi.org/10.1016/j.cosust.2015.12.006

Ostrom, E., 2010a. Beyond Markets and States: Polycentric Governance of Complex Economic Systems. American Economic Review 100, 641–672. https://doi.org/10.1257/aer.100.3.641

Ostrom, E., 2010b. Analyzing collective action. Agricultural Economics 41, 155–166. https://doi.org/10.1111/j.1574-0862.2010.00497.x

Ostrom, E., 1998. A Behavioral Approach to the Rational Choice Theory of Collective Action: Presidential Address, American Political Science Association, 1997. The American Political Science Review 92, 1–22. https://doi.org/10.2307/2585925

Ostrom, E., 1990. Governing the Commons: The Evolution of Institutions for Collective Action. Cambridge University Press.

Pe'er, G., Bonn, A., Bruelheide, H., Dieker, P., Eisenhauer, N., Feindt, P.H., Hagedorn, G., Hansjürgens, B., Herzon, I., Lomba, Â., Marquard, E., Moreira, F., Nitsch, H., Oppermann, R., Perino, A., Röder, N., Schleyer, C., Schindler, S., Wolf, C., Zinngrebe, Y., Lakner, S., 2020. Action needed for the EU Common Agricultural Policy to address sustainability challenges. People and Nature 2, 305–316. https://doi.org/10.1002/pan3.10080

Perrot, T., Gaba, S., Roncoroni, M., Gautier, J.-L., Bretagnolle, V., 2018. Bees increase oilseed rape yield under real field conditions. Agriculture, Ecosystems & Environment 266, 39–48. https://doi.org/10.1016/j.agee.2018.07.020

Perrot, T., Gaba, S., Roncoroni, M., Gautier, J.-L., Saintilan, A., Bretagnolle, V., 2019. Experimental quantification of insect pollination on sunflower yield, reconciling plant and field scale estimates. Basic and Applied Ecology 34, 75–84. https://doi.org/10.1016/j.baae.2018.09.005

Pindyck, R.S., Rubinfeld, D.L., 2018. Microeconomics, Ninth edition, global edition. ed, The Pearson series in economics. Pearson, Harlow, England London; New York Boston San Francisco Toronto Sydney Dubai Singapore Hongkong Tokyo Seoul Taipei New Delhi Cape Town Sao Paulo Mexico City Madrid Amsterdam Munich Paris Milan.

Potts, S.G., Imperatriz-Fonseca, V., Ngo, H.T., Aizen, M.A., Biesmeijer, J.C., Breeze, T.D., Dicks, L.V., Garibaldi, L.A., Hill, R., Settele, J., Vanbergen, A.J., 2016. Safeguarding pollinators and their values to human well-being. Nature 540, 220–229. https://doi.org/10.1038/nature20588

Power, A.G., 2010. Ecosystem services and agriculture: tradeoffs and synergies. Philosophical Transactions of the Royal Society of London B: Biological Sciences 365, 2959–2971. https://doi.org/10.1098/rstb.2010.0143

Prager, K., 2015. Agri-environmental collaboratives for landscape management in Europe. Current Opinion in Environmental Sustainability, Sustainability governance and transformation 12, 59–66. https://doi.org/10.1016/j.cosust.2014.10.009

Prager, K., Reed, M., Scott, A., 2012. Encouraging collaboration for the provision of ecosystem services at a landscape scale—Rethinking agri-environmental payments. Land Use Policy 29, 244–249. https://doi.org/10.1016/j.landusepol.2011.06.012

R Core Team, 2018. R: A language and environment for statistical computing. R Foundation for Statistical Computing, Vienna, Austria.

Requier, F., Odoux, J.-F., Tamic, T., Moreau, N., Henry, M., Decourtye, A., Bretagnolle, V., 2015. Honey bee diet in intensive farmland habitats reveals an unexpectedly high flower richness and a major role of weeds. Ecological Applications 25, 881–890. https://doi.org/10.1890/14-1011.1

Salliou, N., Barnaud, C., Vialatte, A., Monteil, C., 2017. A participatory Bayesian Belief Network approach to explore ambiguity among stakeholders about socio-ecological systems. Environmental Modelling & Software 96, 199–209. https://doi.org/10.1016/j.envsoft.2017.06.050

Scott, J., Marshall, G., 2009. A Dictionary of Sociology. Oxford University Press.

Simoncini, R., Ring, I., Sandström, C., Albert, C., Kasymov, U., Arlettaz, R., 2019. Constraints and opportunities for mainstreaming biodiversity and ecosystem services in the EU's Common Agricultural Policy: Insights from the IPBES assessment for Europe and Central Asia. Land Use Policy 88, 104099. https://doi.org/10.1016/j.landusepol.2019.104099

Stefanic, I., Stefanic, E., Puskadija, Z., Kezic, N., Grgic, Z., 2004. Beekeeping in the Republic of Croatia. Bee World 85, 19–21. https://doi.org/10.1080/0005772X.2004.11099608

Stokstad, E., 2007. The Case of the Empty Hives. Science 316, 970–972. https://doi.org/10.1126/science.316.5827.970

Swinton, S.M., Lupi, F., Robertson, G.P., Hamilton, S.K., 2007. Ecosystem services and agriculture: Cultivating agricultural ecosystems for diverse benefits. Ecological Economics, Special Section - Ecosystem Services and Agriculture 64, 245–252. https://doi.org/10.1016/j.ecolecon.2007.09.020

Vanni, F., 2013. Agriculture and Public Goods: The Role of Collective Action. Springer Science & Business Media.

Vialatte, A., Barnaud, C., Blanco, J., Ouin, A., Choisis, J.-P., Andrieu, E., Sheeren, D., Ladet, S., Deconchat, M., Clément, F., Esquerré, D., Sirami, C., 2019. A conceptual framework for the governance of multiple ecosystem services in agricultural landscapes. Landscape Ecol 34, 1653–1673. https://doi.org/10.1007/s10980-019-00829-4

Williams, I.H., 1994. The dependence of crop production within the European Union on pollination by honey bees [WWW Document]. Evans, K [Editor] Agricultural Zoology Reviews. URL https://geoscience.net/research/002/981/002981332.php (accessed 3.6.18).

Woodcock, B.A., Garratt, M.P.D., Powney, G.D., Shaw, R.F., Osborne, J.L., Soroka, J., Lindström, S. a. M., Stanley, D., Ouvrard, P., Edwards, M.E., Jauker, F., McCracken, M.E., Zou, Y., Potts, S.G., Rundlöf, M., Noriega, J.A., Greenop, A., Smith, H.G., Bommarco, R., Werf, W. van der, Stout, J.C., Steffan-Dewenter, I., Morandin, L., Bullock, J.M., Pywell, R.F., 2019. Meta-analysis reveals that pollinator functional diversity and abundance enhance crop pollination and yield. Nature Communications 10, 1481. https://doi.org/10.1038/s41467-019-09393-6

Zhang, H., Breeze, T., Bailey, A., Garthwaite, D., Harrington, R., Potts, S.G., 2017. Arthropod Pest Control for UK Oilseed Rape – Comparing Insecticide Efficacies, Side Effects and Alternatives. PLoS One 12. https://doi.org/10.1371/journal.pone.0169475

1 # Tables

2 ## Table 1. Functional forms used in the model

| Stakeholder | Function | Equation | Description |
|---|---|---|---|
| Farmer | Wheat benefits | $\eta_W = \chi_W x_W^{\varepsilon}$ (2) | Cobb-Douglas function, exhibiting usual properties of production functions: positive and diminishing marginal productivity of the production input. $\chi_W$ is the gross margin of wheat, $\varepsilon$ is the input elasticity. |
| Farmer | Grassland benefits | $\eta_G = \chi_G x_G^{\varepsilon}$ (3) | Cobb-Douglas function. $\chi_G$ is the gross margin of hay, $\varepsilon$ is the input elasticity. |
| Farmer | OSR production | $F_{OSR} = x_{OSR} \sum_{j=1}^{3} f_{OSR,j}$ (4) | Linear combination of three partial yields $j \in \{1,2,3\}$ (giving total yield), times OSR area (Montoya et al. 2019) |
| Farmer | Bee-dependent crop yield | $f_{OSR,1} = \alpha_1 \dfrac{B}{\beta_1 + B}$ (5) | Saturating type II functional form, describing the saturating uptake of resources (Holling 1973), broadly inspired from the model of Montoya et al. (2019). $0 < B < 1$ is the total bee abundance. $\alpha_1 > 0$ is a parameter related to the level of dependence, and $0 \leq \beta_1 \leq 1$ is a parameter of bee efficacy. |
| Farmer | Pesticide-dependant crop yield | $$f_{OSR,2} = \alpha_2 \frac{x_p'}{\beta_2 + x_p'} \quad (6)$$ | By extension, saturating type II function, describing the saturation of pesticides. Widely used in crop response models (Fernandez-Cornejo, Jans & Smith 1998). $\alpha_2 > 0$ is a parameter related to the level of dependence, and $0 < \beta_2 \leq 1$ is a parameter of pesticide efficacy. |
| Beekeeper | Honey production | $$F_H = f_H x_{OSR}^{\gamma_1} x_H^{\gamma_2} D \quad (7)$$ | Cobb-Douglas function with OSR area and beehives as inputs, following the Siebert's (1980) model. $0 < \gamma_1 < 1$ and $0 < \gamma_2 < 1$ are the partial elasticities of production. $f_H$ is a production constant, $D$ is linked to yield losses caused by pesticides (Eq.7). |
| | Damage function | $$D = 1 - \delta x_{OSR}[x_p]^\nu \quad (8)$$ | Complementary of honey and wild bee losses. $\delta > 0$ and $0 < \nu < 1$ are constant parameters calibrated with recent studies (Chambers, Chatzimichael & Tzouvelekas 2019; Wintermantel *et al.* 2020) |

**Table 2. Bioeconomic indicators**

| Type | Indicator | Expression | Description |
|---|---|---|---|
| Economic | Stakeholders' wealth | $E_1 = \Pi^{(f)} + \Pi^{(k)}$ | Sum of representative beekeepers' and farmers' profits. |
| | Total wealth | $E_2 = E_1 - \tau^{(f)} - \tau^{(k)}$ | Real wealth at the landscape scale, removing transfers from policy maker to beekeepers (subsidies) or from farmers to policy maker (tax). The difference of each policy scenario from BAU represents the additional wealth coming from policy implementation. |
| Ecosystem services | Pollination | $ES_1 = \overline{B}$ | Bee abundance |
| | Food/Feed provision | $ES_2 = \xi_W Y_W x_W + \xi_G Y_G x_G + \xi_{OSR} F_{OSR} + \xi_H F_H$ | Sum of agricultural/apicultural productions, expressed in units of energy. $\xi_i$ are energy multiplicative shifters of commodities. $Y_j$ are average yields of concerned crops. |
| | Water quality (pesticides) | $ES_3 = 1 - x_{OSR} x_p - \theta x_W$ | Opposite of pesticide quantities in OSR ($x_{OSR} x_p$) and wheat ($\theta x_W$). $\theta$ is the toxicity index of |
| | | | wheat relatively to OSR. Grasslands are assumed to be pesticide free. This indicator is related to the ES of water provisioning for drinking (Grizzetti et al. 2016). |
| | Water quality (nutrients) | $ES_4 = 1 - x_{OSR} - x_W$ | Opposite of OSR and wheat areas. This underlies that only these crops are chemically fertilized and they are at the same rate. This indicator is related to the ES of water provisioning for drinking (Grizzetti et al. 2016). |
| Ecological | Wild bee abundance | $BC_1 = x_G k_w D$ | See Eq.9 |
| | Plant species richness | $BC_2 = x_G^s$ | Species-area relationship proxy, a power law with $s$ being the constant slope of the logarithm (Crawley and Harral 2001). |

# Figure legends

**Figure 1. Model overview** The farmers select the amount of pesticide use and land use allocation that maximizes their profits, while beekeepers choose the number of beehives. These decisions directly impact ES provision and bee populations. The latter affect in turn agricultural production through pollination and honey production through honeybee foraging. This creates a mutual interdependency, represented in (b), based on honeybees. Communication as well as incentives can modify the initial choices of stakeholders.

**Figure 2. Consequences of scenarios on economic performance**. Gains and losses without communication (black) and with communication (grey) modelled for each sector (with $n=6$ farmers and $m=1$ beekeeper) and for the policymaker for the three scenarios.

**Figure 3. Bundles of indicators**. Magnitudes of ES (orange), biodiversity conservation indicators (in green) and economic indicators (blue) for each scenario with communication (a) and without (b), shown relative to BAU (0%) (red dashed line). The HS (hive subsidy, light green), PS (price subsidy, dark green) and PT (pesticide tax, blue) scenarios are represented by solid lines.

**Figure 4. Land use in the four scenarios**. Land use in the business as usual (BAU, a), hive subsidy (HS, b), honey price subsidy (PS, c) and pesticide tax (PT, d) scenarios. The cropping mix is composed of wheat (dark yellow), oilseed rape (light yellow) and grasslands (green).

**Figure 5. Decisional cascade and agro-ecological interdependency**. Variation from the BAU scenario in terms of marginal economic and environmental indicators for the subsidy scenarios (a, b, c: HS in light green, PS in dark green) and the tax scenario (d, e, f: PT in blue). The first row (a, d) shows how the marginal revenue of outputs (honey and oilseed rape seeds) is impacted by the policy (a: the beekeeper and d: the farmer). The second row (b, e) illustrates the effects of the management changes on pollination, which is the keystone of the interdependency. The third row (c, f) shows how the marginal product of each productive input for the non-targeted stakeholder is indirectly affected (c: farmer and f: beekeeper).

# Figures

**Figure 1**

(a) Overview of model functioning

(b) Mutual interdependency

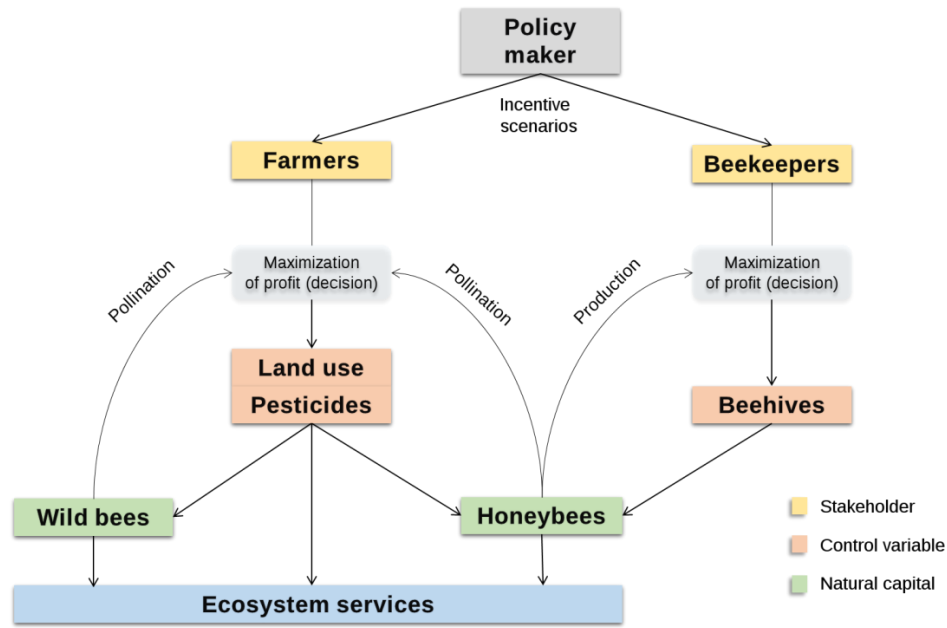
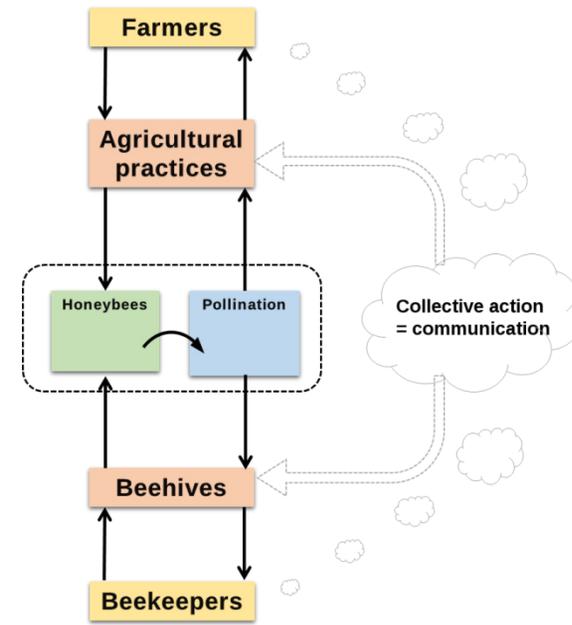

**Figure 2**

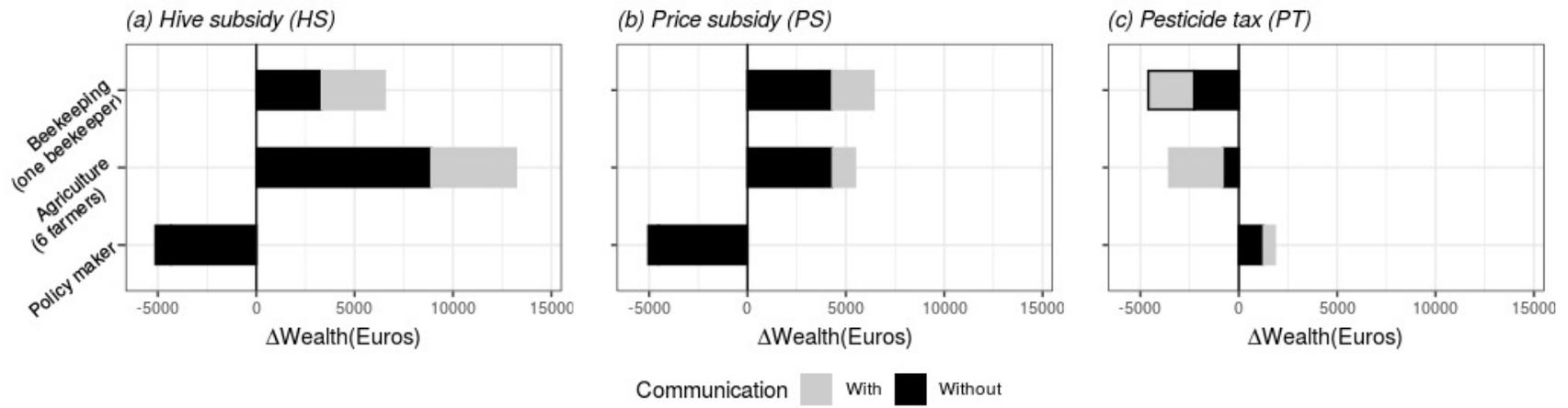

**Figure 3**

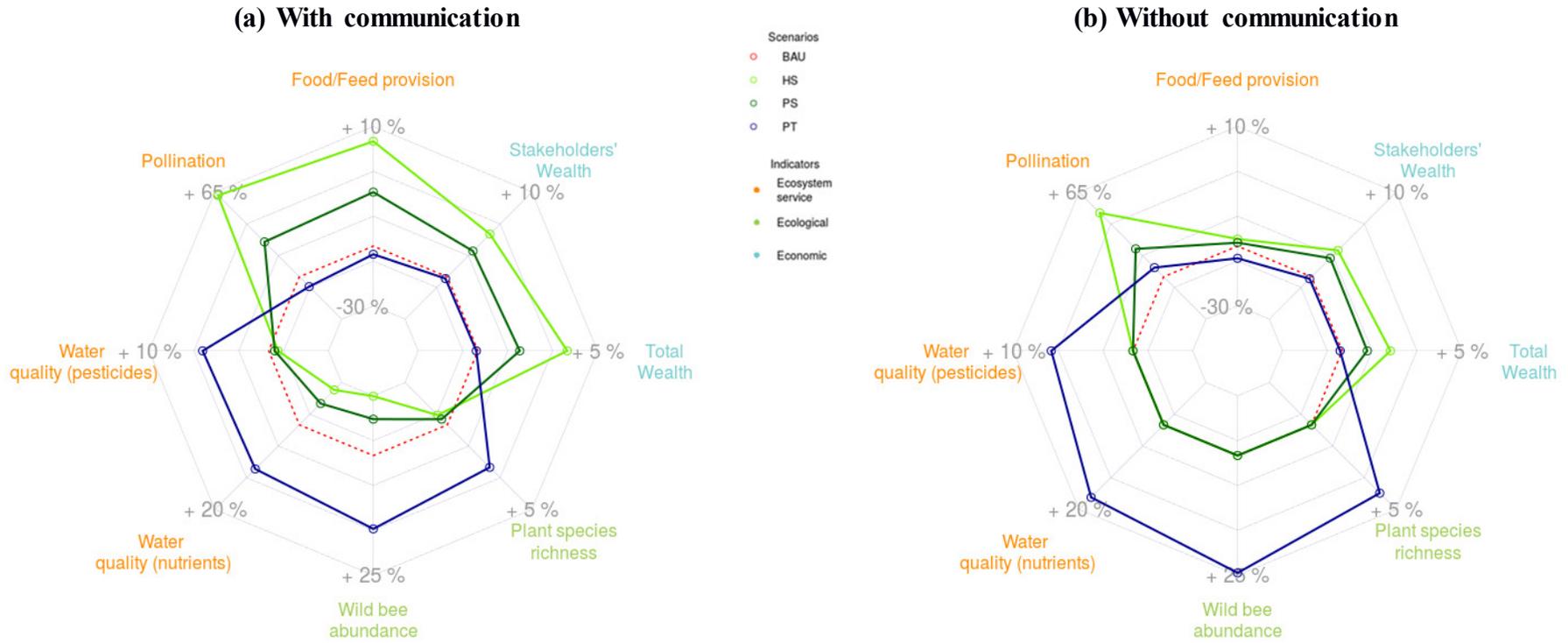

**Figure 4**

Land use in four scenarios

(a) Business as usual (BAU)
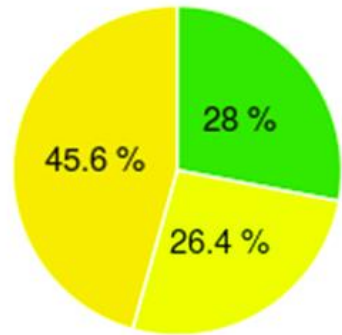

(b) Hive subsidy (HS)
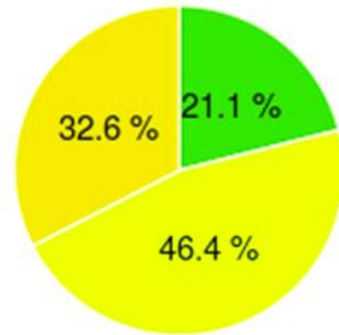

(c) Price subsidy (PS)
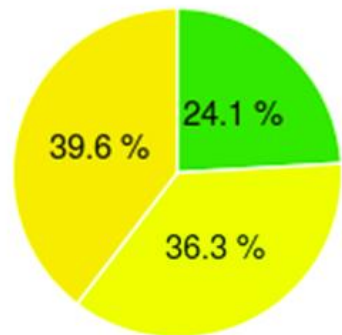

(d) Pesticide tax (PT)
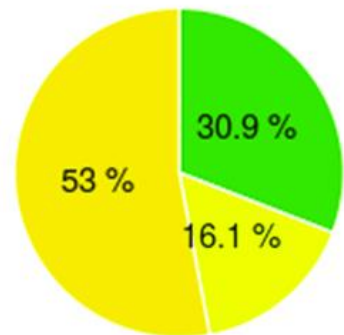

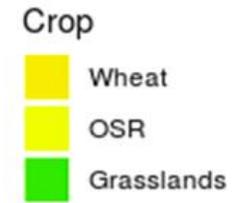

1   **Figure 5**

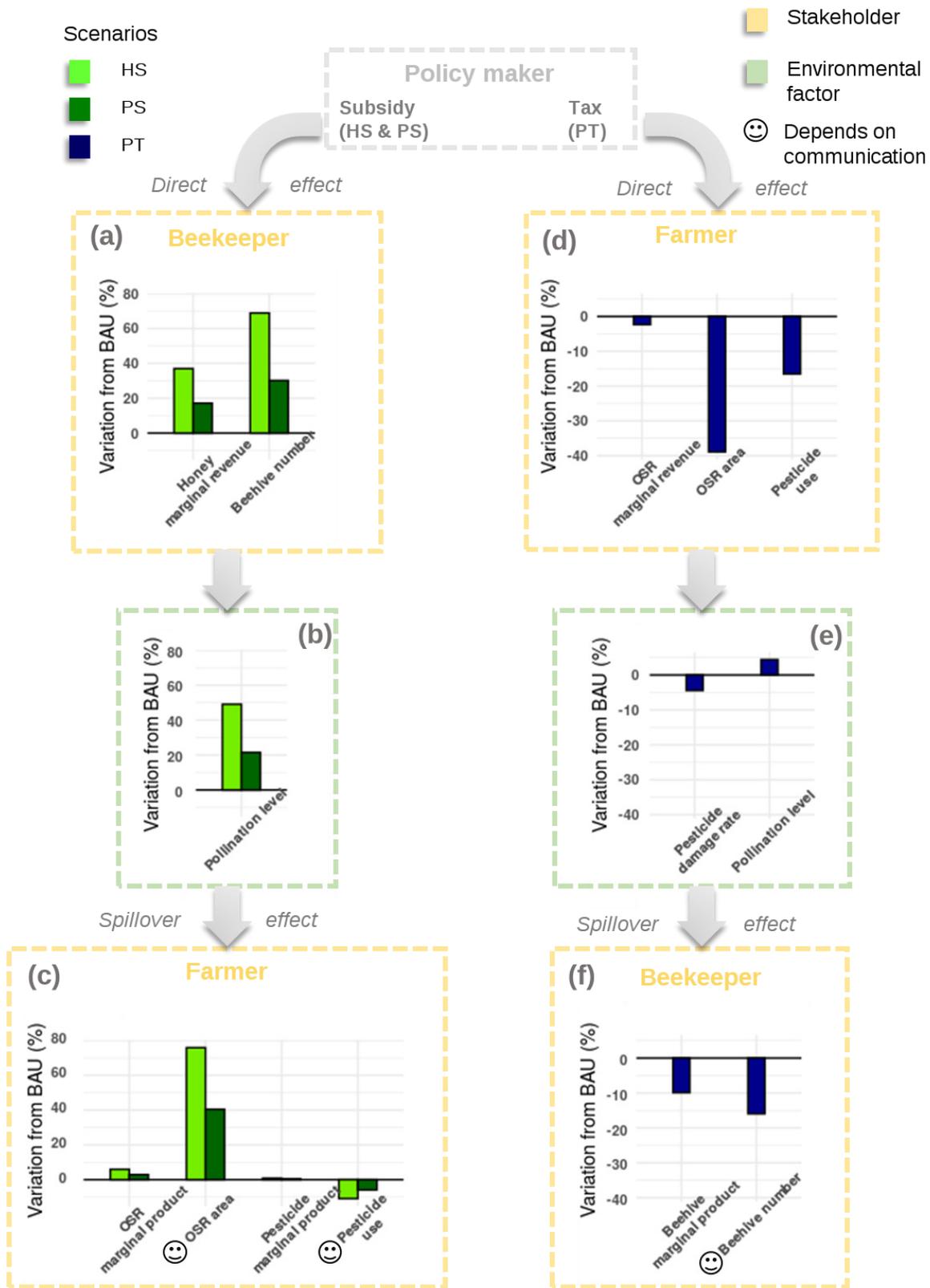

# Appendices

## Appendix S1: Effect of pesticide applications on honeybees

There are evidences that pesticide applications negatively affect honeybees (Johnson et al., 2010, Woodcock et al., 2017). In our model, this negative effect is represented by a damage function $D$ that linearly decreases honey production as follows: $F_H = f_H x_{OSR}^{\gamma_1} x_H^{\gamma_2} D$ (see Eq.7, Table 1). The negative effect of pesticide applications on honeybee is not explicitly but implicitly included in the honeybee population model. The damage function impacts the honeybee population model, through its effects on the optimal number of beehives $x_H^*$ which decreases with pesticides (with Eq.9 : $x_H^* = \frac{c_H}{\gamma_2 f_H x_{OSR}^{\gamma_1} D}^{\frac{1}{\gamma_2 - 1}}$ ). The negative effect of pesticides on honeybees can however be indirectly computed and yields to a negative effect of rate $1/D^{1/1-\gamma_2}$. Because $\gamma_2 < 1$ and $1/D^{1/1-\gamma_2} > D, \forall D \in [0,1]$, honey production is less affected by pesticides than honeybees. This result, which is in accordance with Khoury et al. (2013) and Chambers et al. (2019), also reflects a greater effect of pesticides on honeybees compared to wild bees. This is a reasonable assumption since honey bees are more frequently found in oilseed rape fields compared to wild bees (Rollin et al. 2013).

References

Chambers, R. G., Chatzimichael, K., & Tzouvelekas, V. (2019). Sub-lethal concentrations of neonicotinoid insecticides at the field level affect negatively honey yield: Evidence from a 6-year survey of Greek apiaries. PloS one, 14(4).

Johnson, R. M., Ellis, M. D., Mullin, C. A., & Frazier, M. (2010). Pesticides and honey bee toxicity–USA. *Apidologie*, *41*(3), 312-331.

Khoury, D. S., Barron, A. B., & Myerscough, M. R. (2013). Modelling food and population dynamics in honey bee colonies. PloS one, 8(5).

Rollin, O., Bretagnolle, V., Decourtye, A., Aptel, J., Michel, N., Vaissière, B. E., & Henry, M. (2013). Differences of floral resource use between honey bees and wild bees in an intensive farming system. Agriculture, Ecosystems & Environment, 179, 78-86.

Woodcock, B. A., Bullock, J. M., Shore, R. F., Heard, M. S., Pereira, M. G., Redhead, J., ... & Peyton, J. (2017). Country-specific effects of neonicotinoid pesticides on honey bees and wild bees. Science, 356(6345), 1393-1395.

# Appendix S2: Marginal economic indicators

In order to highlight the decisional cascade created by the implementation of incentive, we defined two microeconomic indicators for each scenario. We selected these indicators following Meade (1952). The optimal provision of the variable $i$ in scenario $k$ is $x^*_{i,k}$ and the revenue and production of agent $i$ in scenario $k$ is $R_k^{(i)}$ and $F_k^{(i)}$ respectively.

<u>Decisional cascade coming from the implementation of HS and PS scenarios</u> (Figure 5a-c).

The subsidies scenarios, HS and PS, aim at motivating beekeepers to increase the number of beehives. This results in an increase of the number of honeybees that positively affects the farmers' management strategy through a spillover effect because of the interdependency between the beekeeper and the farmer due to pollination ES. The spillover effect, defined as the indirect effect on related stakeholders, can be estimated from the difference in the marginal products of OSR-related inputs ($x_{OSR}$ and $x_p$) of the BAU and HS(PS) scenarios. The direct effect of the subsidies scenarios is measured by the marginal revenues of honey production (Table S1a).

**Table S1a**: Mathematical formulations.

|  | BAU scenario | HS scenario | PS scenario |
|---|---|---|---|
| marginal revenues of honey production | $\dfrac{\partial R_{BAU}^{(k)}}{\partial F_H}(x_H^*, x_{OSR}^*, x_p^*)_{BAU}$ | $\dfrac{\partial R_{HS}^{(k)}}{\partial F_H}(x_H^*, x_{OSR}^*, x_p^*)_{BAU}$ | $\dfrac{\partial R_{PS}^{(k)}}{\partial F_H}(x_H^*, x_{OSR}^*, x_p^*)_{BAU}$ |
| marginal products of OSR- | $\dfrac{\partial F_{OSR,BAU}}{\partial x_{OSR}(x_p)}(X^*, x_p^*, x_H^*)_{BAU}$ | $\dfrac{\partial F_{OSR,HS}}{\partial x_{OSR}(x_p)}(X^*, x_p^*, x_H^*)_{BAU}$ | $\dfrac{\partial F_{OSR,PS}}{\partial x_{OSR}(x_p)}(X^*, x_p^*, x_H^*)_{BAU}$ |
| related inputs | | | |

Decisional cascade coming from the implementation of the tax scenario (Figure 5d-f).

The tax scenario (PT) aims at discouraging farmers from using large amount of pesticides. Because of the farmer-beekeeper interdependency, the farmer's decision to reduce pesticide use will impact the beekeeper through a spillover effect (i.e. lower pesticide use will result in a higher honeybee survival rate and higher honey production). The spillover effect can be estimated from the difference in the marginal products of beehives between the BAU and AT scenarios. The direct effect of taxing pesticide use is measured on the farmers' marginal revenues from OSR production. Table S1b presents the mathematical formulations.

**Table S1b.** Mathematical formulation

| | **BAU scenario** | **AT scenario** |
|---|---|---|
| marginal revenues of OSR production | $\frac{\partial R_{BAU}^{(f)}}{\partial F_H}\left(X^*, x_p^*, x_H^*\right)_{BAU}$ | $\frac{\partial R_{PT}^{(f)}}{\partial F_H}\left(X^*, x_p^*, x_H^*\right)_{BAU}$ |
| marginal products of beehives | $\frac{\partial F_{H,BAU}}{\partial x_H}\left(x_H^*, x_{OSR}^*, x_p^*\right)_{BAU}$ | $\frac{\partial F_{H,PT}}{\partial x_H}\left(x_H^*, x_{OSR}^*, x_p^*\right)_{BAU}$ |

References

Meade, J. E. (1952). External economies and diseconomies in a competitive situation. The economic journal, 62(245), 54-67.

# Appendix S3 - Calibration table

| Parameter | Expression | Value | Unit | Reference(s) |
|---|---|---|---|---|
| Insect-pollination dependence | $\alpha_1$ | 1.7 | mass.area$^{-1}$ | Perrot et al. (2018) and Data from LTSER Plaine & Val de Sevre |
| Bee efficiency (half saturation) | $\beta_1$ | 0.3 | dimensionless | Perrot et al. (2018) and Data from LTSER Plaine & Val de Sevre |
| Pesticide-dependence | $\alpha_2$ | 1 | mass.area$^{-1}$ | Fernandez-Cornejo et al. (1998); Zhang et al. (2017) and Data from LTSER Plaine & Val de Sevre |
| Pesticide efficiency (half saturation) | $\beta_2$ | 0.3 | dimensionless | Fernandez-Cornejo et al. (1998); Zhang et al. (2017) and Data from LTSER Plaine & Val de Sevre |
| OSR yield derived from other production inputs | $f_{OSR,3}$ | 1.55 | mass.area$^{-1}$ | Data from LTSER Plaine & Val de Sevre |
| Hay gross margin | $\chi_G$ | 400 | money.area$^{-1}$ | PEREL (2015) |
| Wheat gross margin | $\chi_W$ | 550 | money€.area$^{-1}$ | Cerfrance (2018) |
| Elasticity of agricultural production | $\varepsilon$ | 0.55 | dimensionless | Ory (2020) |
| Honey production constant | $f_H$ | 20 | mass.beehive$^{-1}$ | Bova (2012) |
| Beehive elasticity | $\gamma_1$ | 0.4 | dimensionless | Champetier et al. (2015); Vaziritabar et al. |
| | | | | (2014); Siebert (1980) |
| Flowering crop elasticity | $\gamma_2$ | 0.3 | dimensionless | Champetier et al., (2015); Vaziritabar et al. (2014); Siebert (1980) |
| Maximal number of beehives | $\overline{x_H}$ | 0.72 | unit.area$^{-1}$ | Bova (2012) |
| OSR price per ton | $p_{OSR}$ | 350 | money.mass$^{-1}$ | Visionet (2020) |
| Marginal cost of OSR inputs apart from pesticides | $c_{OSR}$ | 550 | money.area$^{-1}$ | Terres Inovia (2018) |
| Marginal cost of pesticides | $c_p$ | 150 | money.area$^{-1}$ | Terres Inovia (2018) |
| Honey price | $p_H$ | 7 | money.mass$^{-1}$ | Bova (2012) |
| Marginal cost of beehives | $c_H$ | 100 | money.unit$^{-1}$ | Bova (2012) |
| Per hive subsidy | $z_{HS}$ | 28 | money.unit$^{-1}$ | computed |
| Price subsidy | $z_{PS}$ | 1.22 | money.mass$^{-1}$ | computed |
| Pesticide tax | $z_{PT}$ | 50 | money.area$^{-1}$ | Böcker and Finger (2016); Jacquet et al. (2011) |
| Honeybee carrying capacity | $k_H$ | 18900 | animal.unit$^{-1}$ | Bretagnolle and Gaba (2015) |
| Wild bee carrying capacity | $k_w$ | 5000 | animal.area$^{-1}$ | Osborne et al. (2008) |
| Pesticide damage | $\delta$ | 0.5 | dimensionless | Chambers et al. (2019) ; Wintermantel et |
| parameter | | | | al. 2020 |
| Pesticide damage parameter | $\nu$ | 0.5 | dimensionless | Chambers et al. (2019) ; Wintermantel et al. 2020 |
| Slope of the logarithm of species richness against the logarithm of area | $s$ | 0.25 | dimensionless | Crawley and Harral (2001) |
| Energy density of OSR | $\xi_{OSR}$ | 6,940,000 | energy.mass$^{-1}$ | INRA (2019) |
| Energy density of wheat | $\xi_W$ | 4,350,000 | energy.mass$^{-1}$ | INRA (2019) |
| Energy density of grasslands | $\xi_G$ | 3,417,000 | energy.mass$^{-1}$ | INRA (2019) |
| Energy density of honey | $\xi_H$ | 3,270 | energy.mass$^{-1}$ | |
| Wheat-OSR toxicity relative index | $\theta$ | 0.285 | dimensionless | Data from LTSER Plaine & Val de Sevre |

References

Böcker, T., Finger, R., 2016. European Pesticide Tax Schemes in Comparison: An Analysis of Experiences and Developments. Sustainability 8, 378. https://doi.org/10.3390/su8040378

Bova, F., 2012. Audit économique de la filière apicole française, Les synthèses de FranceAgriMer. FranceAgriMer.

Bretagnolle, V., Gaba, S., 2015. Weeds for bees? A review. Agronomy for Sustainable Development 35, 891–909. https://doi.org/10.1007/s13593-015-0302-5

Cerfrance, 2018. Résultats economiques 2017.

Chambers, R.G., Chatzimichael, K., Tzouvelekas, V., 2019. Sub-lethal concentrations of neonicotinoid insecticides at the field level affect negatively honey yield: Evidence from a 6-year survey of Greek apiaries. PLoS One 14. https://doi.org/10.1371/journal.pone.0215363

Champetier, A., Sumner, D.A., Wilen, J.E., 2015. The Bioeconomics of Honey Bees and Pollination. Environ Resource Econ 60, 143–164. https://doi.org/10.1007/s10640-014-9761-4

Crawley, M.J., Harral, J.E., 2001. Scale Dependence in Plant Biodiversity. Science 291, 864–868. https://doi.org/10.1126/science.291.5505.864

Fernandez-Cornejo, J., Jans, S., Smith, M., 1998. Issues in the Economics of Pesticide Use in Agriculture: A Review of the Empirical Evidence. Applied Economic Perspectives and Policy 20, 462–488. https://doi.org/10.2307/1350002

INRA, CIRAD, AFZ, 2019. Table par catégorie (valeurs sur matière sèche) [WWW Document]. Feetables. URL https://feedtables.com/fr/content/table-dry-matter (accessed 7.8.19).

Jacquet, F., Butault, J.-P., Guichard, L., 2011. An economic analysis of the possibility of reducing pesticides in French field crops. Ecological Economics, Special Section - Governing the Commons: Learning from Field and Laboratory Experiments 70, 1638–1648. https://doi.org/10.1016/j.ecolecon.2011.04.003

Ory, X., 2020. Lien entre la taille des exploitation agricoles, leur productivite et leur impact sur l'environnement (No. 2020/2), Documents de travail. Ministere de l'economie et des finances.

Osborne, J.L., Martin, A.P., Shortall, C.R., Todd, A.D., Goulson, D., Knight, M.E., Hale, R.J., Sanderson, R.A., 2008. Quantifying and comparing bumblebee nest densities in gardens and countryside habitats. Journal of Applied Ecology 45, 784–792. https://doi.org/10.1111/j.1365-2664.2007.01359.x

PEREL, 2015. Fiche synthèse sur les coûts des fourrages rendus à l'auge : Prairie naturelle.

Perrot, T., Gaba, S., Roncoroni, M., Gautier, J.-L., Bretagnolle, V., 2018. Bees increase oilseed rape yield under real field conditions. Agriculture, Ecosystems & Environment 266, 39–48. https://doi.org/10.1016/j.agee.2018.07.020

Siebert, J.W., 1980. Beekeeping, Pollination, and Externalities in California Agriculture. American Journal of Agricultural Economics 62, 165–171. https://doi.org/10.2307/1239682

Terres Inovia, 2018. Observatoire des résultats économiques à la production - Analyse 2018, Observatoire des résultats économiques à la productio. Terres Inovia.

Vaziritabar, S., Oshidari, S., Aghamirkarimi, A., 2014. Estimation of honey production function and productivity of its factors in the Alborz province of Iran. Journal of Biodiversity and Environmental Sciences (JBES) 5, 526–533.

Visionet, 2020. Historique des prix trimestriels moyens payés aux producteurs depuis 2005 - cereales et proteagineux.

Wintermantel, D., Odoux, J.-F., Decourtye, A., Henry, M., Allier, F., Bretagnolle, V., 2020. Neonicotinoid-induced mortality risk for bees foraging on oilseed rape nectar persists despite EU moratorium. Science of The Total Environment 704, 135400. https://doi.org/10.1016/j.scitotenv.2019.135400

Zhang, H., Breeze, T., Bailey, A., Garthwaite, D., Harrington, R., Potts, S.G., 2017. Arthropod Pest Control for UK Oilseed Rape – Comparing Insecticide Efficacies, Side Effects and Alternatives. PLOS ONE 12, e0169475. https://doi.org/10.1371/journal.pone.0169475

# Appendix S4 - Sensitivity analysis on oilseed rape price

We performed a sensitivity analysis on oilseed rape (OSR) price based on market price past values (2013-2017). During these years the OSR price for French farmers ranged approximately from 300 to 400€/t (Terres Inovia 2018).

| OSR price | Land use |
|---|---|
| 300€/t | 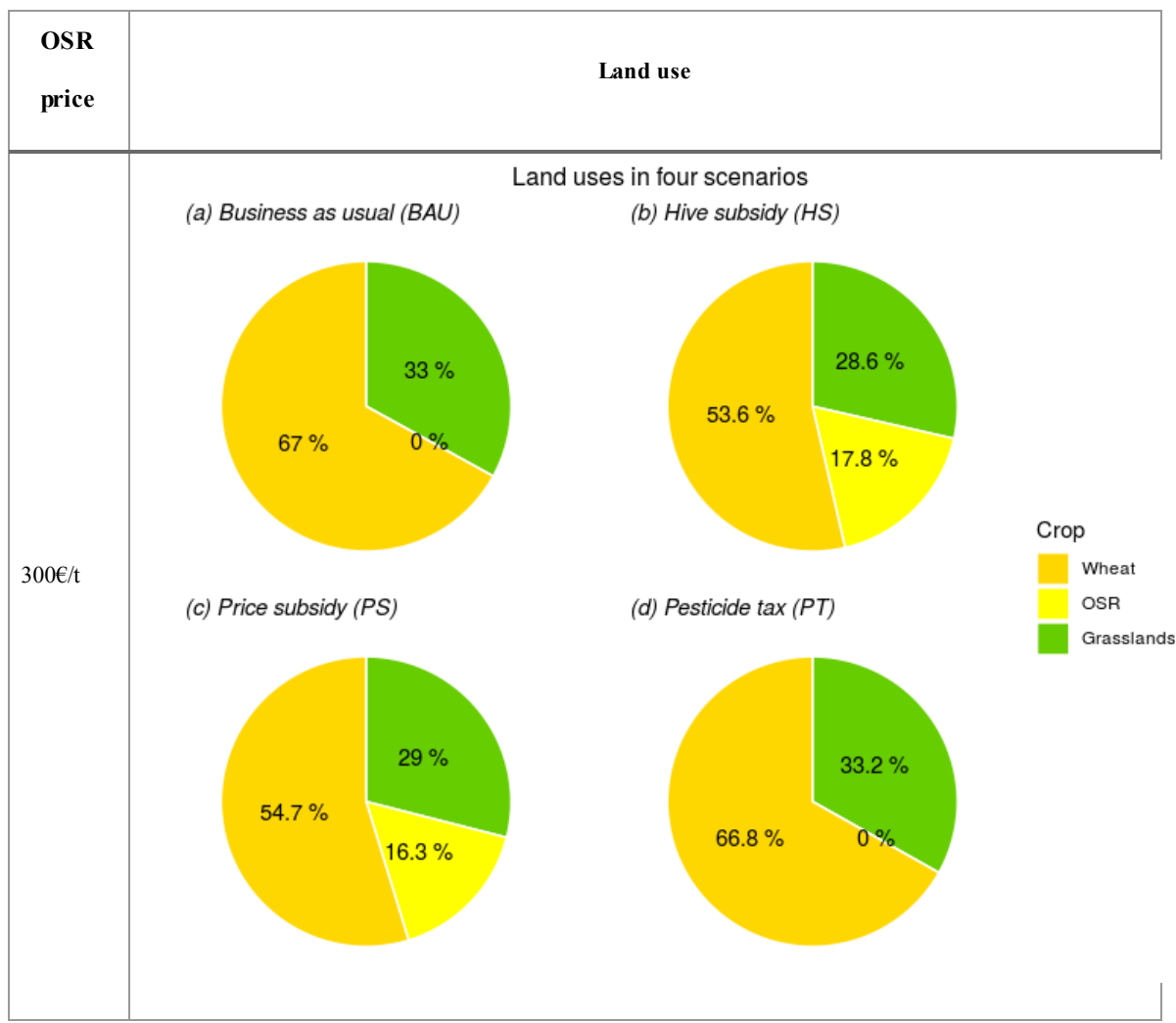 |

400€/t

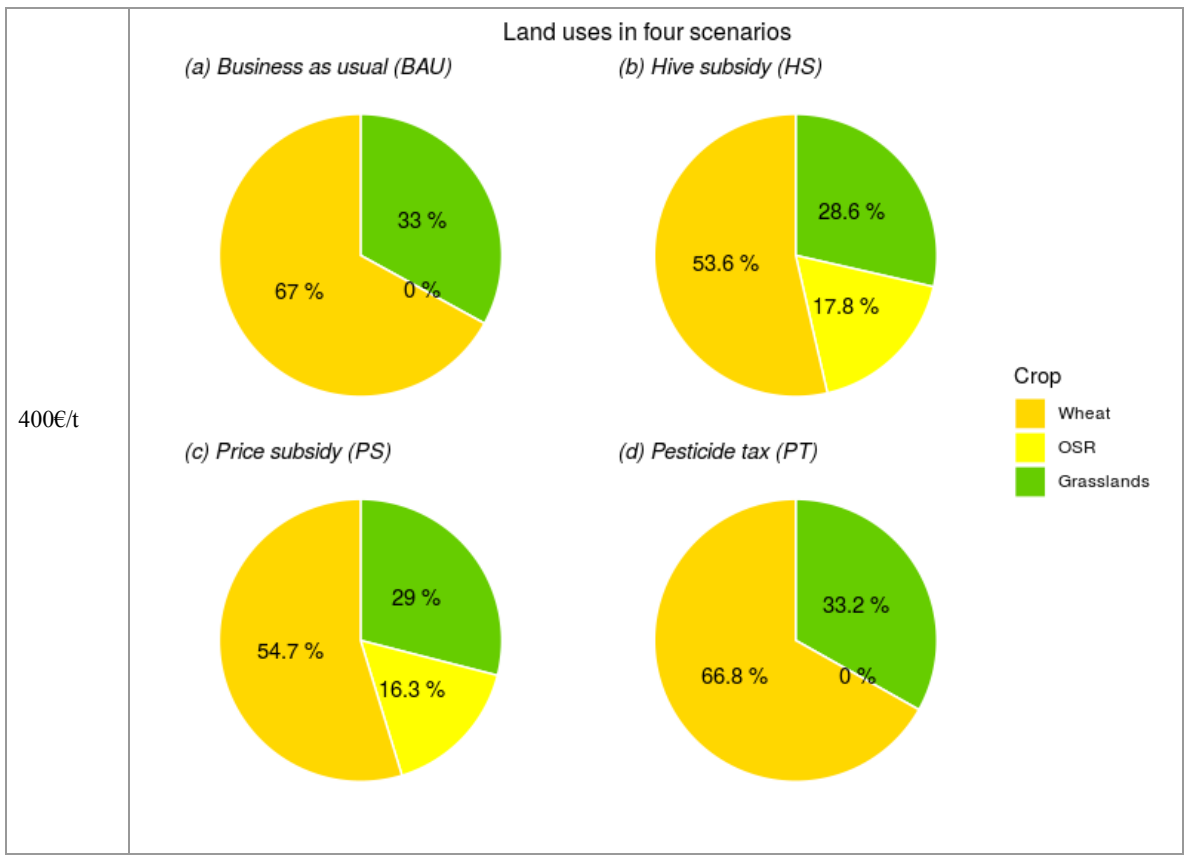

We observed that land use is widely dependent on OSR price. A lower price logically leads to less OSR area : with our calibration OSR was no longer even grown by farmers in BAU. Conversely, a higher price leads to more OSR area. These changes in OSR area are accompanied with higher (if OSR area decrease) or lower (if OSR area decrease) other crop areas.